\documentclass[aps,prd,amssymb,showpacs,floatfix,twocolumn,superscriptaddress,nofootinbib]{revtex4-2}
\usepackage{amsmath,amsfonts,graphics,epsfig}

\begin{document}

\title{A step towards estimation of the neutral-hadron size: the
gravitational mass radius of $\pi^0$ meson in a relativistic theory of
composite particles}

\author{A.F.~Krutov}
\email{a$_$krutov@rambler.ru}, \affiliation{Samara State Technical
University, 443100 Samara, Russia} \author{V.E.~Troitsky}
\email{troitsky@theory.sinp.msu.ru} \affiliation{D.V.~Skobeltsyn Institute
        of Nuclear Physics,\\
M.V.~Lomonosov Moscow State University, Moscow 119991, Russia}
\date{\today}

\begin{abstract}
We extend our nonperturbative essentially relativistic approach,
elaborated previously, to perform an approximate estimation 
of the size of $\pi^{0}$
meson. We present detailed argumentation for choosing the mass
mean square radius (MSR) for this purpose. Its value calculated in our
approach, using three model quark-antiquark wave functions in pion, is
$(0.5-0.53)$ fm. The neutral-pion electromagnetic form factor is zero in
accordance with the charge-conjugation symmetry. We demonstrate ambiguities
encountered when using standard definitions of mechanical MSR for the
size. We discuss MSRs obtained in various approaches and their comparison
with each other and with our results.
\end{abstract}

\maketitle

\section{Introduction}
Since ancient times, people were interested in the structure of matter on
scales appropriate to corresponding stages of knowledge.
The gravitational form factors (GFFs) of hadrons are in the focus of
investigation during last decades. The current state of the problem is
reviewed in \cite{PoS18, BuE24}. GFFs encode key information including the
mechanical properties of the particle as well as the conditions of its
stability.

To describe the internal electromagnetic structure of particles,
the electromagnetic form factors (EFFs) of hadrons have been studied
(beginning from 1950s) using the electron as a probing particle. EFFs
describe the distributions of charge, magnetic and quadrupole moments
inside the hadron. Since the gravitational interaction is very weak, the
analogous measurements of GFFs  cannot be carried out in experiments.
However, the information about GFFs can be extracted from the
hard-exclusive processes described in terms of unpolarized generalized
parton distribution (GPD) \cite{PoS18, BuE24}. GFFs contain the
information about the distribution of different
physical characteristics, including spin, mass and mechanical stresses
inside particles.

It is interesting to compare the information about hadron structure
obtained through EFFs and GFFs, in particular, the distribution of charge
and of mass, and to estimate the relation between the sizes of
hadron extracted from EFFs and from GFFs.
This problem is largely discussed in the literature: see, for example, the
impressive papers \cite{BuE23, BuE24, Mil19, Jaf21} on proton and neutron
gravitational radii (and also recent work \cite{YaX24}, which contains the
most recent calculation of gravitational nucleon radii and work \cite{Pet24},
in which the relationship between nucleon form factors and geometric
dimensions of the nucleon is considered).
Here, we consider this interesting problem from a slightly different point
of view.

In the present paper, we
consider electromagnetic and gravitational structures of neutral and
charged pions.
We use our version
of the relativistic composite-particle theory in the instant-form impulse
approximation approach. Our approach has demonstrated reasonable results
for pion gravitational structure \cite{KrT21, KrT22, KrT23}. The approach
works well in the case of EFFs: back in 2001, we predicted \cite{KrT01},
with surprising accuracy, all the results for charged  pion
EFF obtained later in JLab experiments (see the discussion in
\cite{KrT09prc}). Other approaches have made use of RQM for calculation
of GFFs, see e.g.\ the recent work \cite{CaX24}, where GFFs are
calculated in the light-cone potential approach.
It is worth emphasizing that we calculate GFFs and EFFs in the framework of
a common formalism, using the same approximations and the same model
parameters.

One of the main goals of the present paper is the estimation, at least an
indirect assessment, of the size of the neutral pion. At the same time it
should be taken into account that there is no strict definition of the
size concept.
We show  that, in our approach, the charge form factor of neutral pion is
zero in accordance with
theoretical expectations.
The
experimental value of the slope of this form factor at zero, that is
the charge MSR, vanishes within the measurement errors
\cite{Kho20, Beh91, Dre92, Far92, Nav24}. So, the charge MSR of $\pi^0$
cannot be used to obtain its size.

There are also serious reasons not to consider the mechanical MSR as a
reliable value to estimate the pion size. They are connected with the
calculation of the integral over the full region of momentum transfers and
are discussed in Section \ref{sec: Sec 2}.
So, we choose the mass MSR to obtain an approximate estimation 
of the neutral-pion
size. The mass MSR is defined by the derivative of the pion GFF $A(Q^2)$ at
zero value of the square of momentum transfer.  In our approach, all the
parameters needed to calculate the mass MSR were fixed in previous papers
on the electroweak structure of pion.
We use different model wave functions of quarks in pion and obtain the
results close to one another.

Let us note that even for a more general task of calculating GFFs one needs
only one new parameter,
the $D$-term of the constituent quark. We fix it by fitting the slope of
the pion form factor $D$ at zero to the value extracted in \cite{KuS18}
from experimental data.

The results we obtain here are compared with those of other authors,
in particular, with
mass MSR, extracted from the process of creation of two pions
$\gamma^*\gamma\to\pi^0\pi^0$ \cite{KuS18}, with the calculation in
the Nambu-Jona-Lasinio model \cite{FrC19}, with QCD sum rules \cite{AlB21},
in lattice QCD  \cite{BrD05, ShD19, PeH22, HaO23, HaP24}, in holographic
light-front QCD \cite{LiV24, FuI24}, in meson dominance model \cite{BrA24},
in contact interaction approach \cite{SuX24} and in continuum Schwinger
function method with rainbow-ladder truncation \cite{XuD24}.
Possible reasons for differences in results are discussed.

The rest of the paper is organized as follows.
In Sect.\ref{sec: Sec 2} we discuss different MSR radii of particles as
possible quantities to define the size of neutral pion. In
Sect.\ref{sec: Sec 3} we present our relativistic approach and
derive the explicit formulae for pion EFF and GFFs in its framework.
In Sect.\ref{sec: Sec 4} we calculate the pion MSRs and compare our results
with those of other approaches. We briefly
conclude and discuss the perspective in Sect.\ref{sec: Sec 5}.

\section{Charge, mass, and mechanical radii of a particle}
\label{sec: Sec 2}
Let us discuss the properties of different quantities used to characterize
the size of a particle and the limitations that arise in the corresponding
approaches.

\subsection{Charge radius}
\label{sec:Sec2:charge}
The size of a charged particle was initially defined
as its charge radius (MSR), which for the pion is
(see, e.g., \cite{Gou74}):
\begin{equation}
\langle r^2\rangle_{\rm charge} = -6\,\left.\frac{dF^{(\pi)}}{dQ^2}\right|_{Q^2=0}\;,
\label{R2ch}
\end{equation}
where $F^{(\pi)}\equiv F^{(\pi)}(Q^2)$ is the pion EFF, $\,Q^2=-t$, and
$\,t$ is the momentum-transfer squared; $F^{(\pi)}(0) = 1$.

The nonrelativistic charge form factor is the Fourier-transform of the
charge density, and Eq.~(\ref{R2ch}) gives the standard charge MSR in the
coordinate representation.
The works \cite{Mil19, Jaf21, Pet24} contain criticism
of such a geometric interpretation of the form factor slope at zero.
In our opinion, this criticism is related to the long-known problem of
interpreting form factors as Fourier transforms of the corresponding
densities of physical quantities (charge, mass, magnetic moment, etc.)
in relativistic theory, as discussed in \cite{Jaf21}. However, quantities of
type (\ref{R2ch}) are experimentally measured with high accuracy and are very
sensitive to the hadron model, so they are calculated in all approaches.
We will adhere to the terminology dating back to the nonrelativistic theory,
where the interpretation of form factors and their connection with geometric
dimensions (\ref{R2ch}) is not in doubt.
The charge MSR can be obtained directly from
the slope of the charge form factor in the vicinity of the zero momentum
transfer, measured, for example, in processes of pion scattering on outer
electron shells of atoms \cite{Ame84, VoT84}. So it is natural to use
(\ref{R2ch}) to estimate the size of a charged particle in electromagnetic
processes.

Similar reasoning are not valid in the case of neutral pion.
In this case the reaction
$\pi^0\to e^+\,e^-\,\gamma$ gives zero for the  charge MSR
(\ref{R2ch})  within measurement errors \cite{Dre92}. This fits
the general statement \cite{Beh91, Dre92} that
single-photon coupling is forbidden by charge-conjugation ($C$) invariance.
From the $C$ symmetry, the elastic charge form factor of $\pi^0$ vanishes.

At first glance it seems that in general, the long-range and
stronger electromagnetic
interaction must result in the charge radius exceeding other radii,
thus denoting the largest value of the particle size. This result was
obtained using empirical estimations in  \cite{XuR23}.  However, it is
difficult to understand that the size of a neutral particle is defined by
the size of its charge distribution. In particular, for the neutron
``charge MSR,'' the experiment indicates a negative value \cite{Nav24}.

\subsection{Mass radius}
\label{sec:Sec2:mass}
To calculate the particle size in terms of GFFs, different MSRs are used
\cite{PoS18, BuE24, KuS18}. Analogously to the charge radius, the mass MSR
of pion is defined as follows,
\begin{equation}
        \langle r^2\rangle_{\rm mass} =
-6\,\frac{1}{A^{(\pi)}(Q^2)}
\left.\frac{dA^{(\pi)}}{dQ^2}\right|_{Q^2=0}\;,
\label{R2mass}
\end{equation}
where $A^{(\pi)}\equiv A^{(\pi)}(Q^2)$ is one of two pion GFFs,
$A^{(\pi)}(0) = 1$.

The $A(Q^2)$ form factor describes the distribution of rest mass 
in pion
and in the nonrelativistic case is just the Fourier-transform of the
density of matter (compare with (\ref{R2ch})). In relativistic approaches,
the relationship between the form factors and corresponding 
density functions
in (\ref{R2ch}), (\ref{R2mass}) is not so obvious. Still, these formulas
are used to estimate the pion size
\cite{PoS18, KuS18, FrC19, AlB21, HaO23, LiV24, SuX24, XuD24}.
The equality (\ref{R2mass}) should be regarded as defining a 
quantity that can be measured experimentally and which gives an 
approximate estimate of the particle size.

More directly related to the geometric dimensions of the pion 
is the MSR of the spatial distribution of the energy density 
(so-called energy MSR),  which is described by the component 
${\cal T}_{00}(\vec r)$ of the energy-momentum tensor 
(see, for example, \cite{PoS18, YaX24, LoM19, Lor20, ChL24}). 
The geometric size of the particle, determined through the 
energy MSR, depend of course on the choice of the reference 
system (see, for example, \cite{Lor20}). 
The most adequate reference systems for the geometric 
interpretation of the energy radius are listed in \cite{LoM19}.
In literature the Breit system is most often used where 
calculations formally yield the same results as in the 
non-relativistic domain (see, for example, \cite{Lor20}). 
In particular, in the Breit system, the energy MSR of the pion 
has the form \cite{ChL24}:
$$
	\langle r^2\rangle_{\rm energy} =
	-6\,\frac{1}{A^{(\pi)}(Q^2)}
	\left.\frac{dA^{(\pi)}}{dQ^2}\right|_{Q^2=0} - 
$$	
\begin{equation}	
	-\frac{3\,D^{(\pi)}}{2\,m_\pi}
		-\frac{3}{4\,m_\pi}\;,
	\label{R2energy}
\end{equation}
where $D^{(\pi)} = D^{(\pi)}(0)$ is
the pion $D$-term $D^{(\pi)}(Q^2)$ is the pion GFF $D$, 
$m_\pi$ is pion mass.

It is possible to estimate approximately the particle size using the mass MSR
(\ref{R2mass}). Its values for proton and neutron 
should be close to each other, as well as for charged
and neutral pions. The mass MSR is the main focus of our study 
in this paper, although we also calculate the energy MSR here.

\subsection{Mechanical radius}
\label{sec:Sec2:mech}
In addition to the charge and mass MSRs,
the mechanical MSR, connected with the gravitational form factor $D$, is
widely used. The most popular definition of the pion mechanical MSR
\cite{BuE24} is
\begin{equation}
        \langle r^2\rangle_{\rm mech} =
6\frac{D^{(\pi)}(0)}{\int_{0}^{\infty}\,D^{(\pi)}(Q^2)dQ^2}
	= \frac{\int d^3r\,r^2\,p_n(r)}{\int d^3r\,p_n(r)}\;,
\label{R2mech1}
\end{equation}
where $p_n(r)$ is the distribution of the pressure caused by
normal, that is directed along radius-vector, internal forces in pion.

The nonrelativistic limit of the
definition (\ref{R2mech1}) has obvious physical meaning: it
defines the characteristic
size of the distribution of internal stresses in a particle \cite{PoS18}.
Indeed, the positive definiteness of the force normal to the surface is
the ideal quantity to measure the size of a particle. When this force
vanishes the particle ceases to be stable.

Now we present some reasons why we believe that using the mechanical radius
is not the best approach for the size estimation. To obtain the
mechanical MSR (\ref{R2mech1}) from experimental data or from theoretical
study one requires information about the behavior of the form factor $D$
in the entire region of momentum transfers. The corresponding measurements
cannot be carried out in experiments neither today, nor in the foreseeable
future. Even the planned experiments with greater kinematic areas and
enlarged  scale of momentum transfers \cite{Akh19-20, And21, Geo22, Chr22,
Bur23} will not allow to estimate the integral in (\ref{R2mech1}) with
sufficient accuracy. We see one more objection against using
(\ref{R2mech1}). Since the integral includes asymptotically large momentum
transfers, it is suitable to exploit the results of perturbative quantum
chromodynamics (QCD) for particle GFFs \cite{KrT23, Tan18, ToM21, ToM22}.
In the case of pions,
there exists a rigorous model-independent
estimation of perturbative QCD (see, e.g., \cite{KrT23, Tan18, ToM21,
ToM22}) at $Q^2\to\infty$: $D^{(\pi)}(Q^2)\sim 1/Q^2$. This estimation is
valid for any behavior of $D^{(\pi)}(Q^2)$ at finite $Q^2$ and gives the
divergence of the integral in the denominator in (\ref{R2mech1}) and the
zero value of the pion mechanical radius.
This is in contrast with the case of a nucleon, for which QCD gives the
behavior $\sim 1/Q^4$ at $Q^2\to\infty$, which ensures the convergence of
the integral in (\ref{R2mech1}) at the upper bound, and so the finite
value of the MSR.

This fact is probably the reason
for exploiting another definition of the
mechanical MSR
(see, e.g., \cite{KuS18, LiV24, SuX24}). It is
constructed similarly to (\ref{R2ch}) and
(\ref{R2mass}) and represents the slope $S^{(\pi)}_D$ of the form factor
$D$ at zero momentum transfer,
\begin{equation}
        \langle \tilde{r}^2\rangle_{\rm mech} = \left(S^{(\pi)}_D\right)^2 =
        -6\,\frac{1}{D^{(\pi)}(Q^2)}\left.\frac{dD^{(\pi)}}{dQ^2}\right|_{Q^2=0}\;.
        \label{R2mech2}
\end{equation}
Note, that unlike (\ref{R2mech1}), the definition (\ref{R2mech2}), also
used sometimes for mechanical MSR, cannot be directly interpreted in terms
of the spatial distribution of mechanical stresses inside the
particle even in a nonrelativistic case.
Its physics
is not obvious \cite{PoS18}. Apparently, this MSR is to be considered as a
parameter describing the linear approximation of GFF at small values of
momentum-transfer square, connected with standard characteristics of
internal stresses in some complicated way. So, the possibility of adequate
use of (\ref{R2mech1}), (\ref{R2mech2}) for estimation of pion mechanical
MSR needs more study. Note that the authors of the work Ref. \cite{PoS18},
in which (\ref{R2mech1}) was derived, admit the insufficient rigor of the
derivation. This point is out of the scope of the present paper.

Let us pay attention to one more approach to calculating mechanical MSR
(\ref{R2mech1}) that is based on the results obtained either in
experiments, or in lattice model calculations, for the pion form factor
$D^{(\pi)}$ for small and intermediate momentum transfer squares (see,
particularly, \cite{BuE23, HaO23, HaP24}). The authors use these values
for fitting a multipole form of the pion form factor
$D^{(\pi)}=D^{(\pi)}(Q^2)$,
\begin{equation}
              D^{\pi}(Q^{2}) = \alpha  /(1 + Q^{2}/\Lambda ^{2})^{n}\;,
\label{Ddip}
\end{equation}
where $\alpha,\;\Lambda,\;n$ are fitting parameters,
and then calculate the integral in
(\ref{R2mech1}).

The finite interval of the values that can be used for the fitting is,
unfortunately, not sufficient enough to consider the result as a reliable
one. Similarly, the analogous calculation of the slope of the form factor
(\ref{R2mech2}) is unsatisfactory because there are too few points near
zero (see discussion in Sec.~\ref{sec: Sec 4}).

So, our definite opinion is that the best way to estimate the size of
hadrons is to use their charge MSR and mass MSR. Wherein the charge
radius of charged particles may be larger than the mass radius, as in the
case of proton \cite{YaX24,HaP24}.  For neutral
pion, only the mass MSR remains.

\section{Instant form relativistic modified impulse
approximation and the calculation of the pion GFFs and EFF}
\label{sec: Sec 3}

Let us recall briefly our approach.
In the calculations we use a particular variant of the
instant-form (IF) Dirac
relativistic quantum mechanics (RQM) \cite{Dir49}
(see also \cite{LeS78, KeP91, Coe92, KrT09, Pol23}).
The approach was successfully used to describe the electromagnetic
structure of composite particles, including pion
\cite{KrT02, KrT03, KrP16, KrT05}. Recently we have shown
that the pion GFFs
can be derived in the same formalism using the same approximations and, in
fact, the same values of the model parameters \cite{KrT21, KrT22, KrT23}.
The only new parameter, the $D$-term of
the constituent quark, is fixed by fitting the slope (\ref{R2mech2}) to
that extracted in the  work  \cite{KuS18}. The three our papers on GFFs
\cite{KrT21, KrT22, KrT23} are extensively used below.

It is important that in RQM, the interaction is taken into account by
including the interaction operator in the algebra of generators of the
Poincar\'e group. We use a version of the instant form of RQM and
include the interaction in the composite
system by adding the interaction operator to the operator of the mass of
the free constituent system by analogy with the conventional IF of RQM.
Note that it is possible to include the interaction in our approach by the
solutions of the Muskhelishvili-Omn\`es-type equations \cite{TrS69, AnS87}.
These solutions represent wave functions of constituent quarks.
The constituents are supposed to lie on the mass shell  and the wave
function of interacting particles is defined as the eigenfunction of the
complete set of observables. In IF RQM, this set of commuting operators is:
\begin{equation}
	{\hat M}_I^2\;
	(\hbox{or}\;\hat M_I = \hat M_0 + \hat V)\;,\quad
	{\hat J}^2\;,\quad \hat J_3\;,\quad \hat {\vec P}\;,
	\label{complete}
\end{equation}
where $\hat M_0$ is the mass operator for the system of particles without interaction,
${\hat V}$ is the operator of interaction, $\hat M_I$ is the mass operator for the
system with interaction, ${\hat J}^2$ is the operator of the square of the
total angular momentum, $\hat J_3$ is the operator of the projection of the total
angular momentum on the $z$ axis and $\hat {\vec P}$ is the operator of the
total momentum. The components of the operator $\hat{\vec J}$ are 
constructed as invariants from the components of the well-known 4-pseudovector 
Pauli-Lubanski operator in a standard way (see, e.g., \cite{Nov76}).

In the IF RQM, the operators ${\hat J}^2\;,\; \hat J_3\;,\; \hat {\vec P}$
coincide with corresponding operators for the composite system without
interaction, and only the term $\hat M_I^2\;(\hat M_I) $
is interaction dependent.
To solve the problem on eigenfunctions of the set (\ref{complete}), it is
necessary to choose a suitable basis in the Hilbert space of the composite
system states (see details in \cite{KrT21}). In the case of a system of
two constituents one can use the two-particle basis of individual spins
and momenta:
\begin{equation}
	|\,\vec p_1\,,m_1;\,\vec p_2\,,m_2\,\!\rangle =
	|\,\vec p_1\,,m_1\,\! \rangle \otimes
	|\, \vec p_2\,,m_2\, \rangle\;,
	\label{p1p2}
\end{equation}
where $\vec p_1,\,\,\vec p_2$ are the 3-momenta of particles,
$m_1,\,m_2$ are the projections of spins to the $z$ axis.

However it is more suitable to use the basis in which the
motion of the center of mass of two particles is separated:
\begin{equation}
	|\,\vec P,\;\sqrt {s},\;J,\;l,\;S,\;m_J\,\rangle\;,
	\label{Pk}
\end{equation}
where $P_\mu = (p_1 +p_2)_\mu$, $P^2_\mu = s$, $\sqrt {s}$
is the invariant mass of the system of two
particles, $l$ is the orbital momentum in the center-of-mass frames
(c.m.s.) of the system, $\vec S\,^2=(\vec S_1 + \vec S_2)^2 =
S(S+1)\;,\;S$ is the total spin in c.m.s., $J$ is the total angular
momentum, $m_J$ is the projection of the total angular momentum.

The bases (\ref{p1p2}) and (\ref{Pk}) are linked
by the Clebsch-Gordan decomposition of a direct
product (\ref{p1p2}) of two irreducible representations of the
Poincar\'e
group into irreducible representations (\ref{Pk}) \cite{KrT21, KrT09}.
The formulas for the corresponding Clebsch-Gordan coefficients are
given in \cite{KrT21, KrT23}.

In the basis (\ref{Pk}), only the operator $\hat M_I$ in the complete
set (\ref{complete}) is non-diagonal. So, the wave function of the
internal motion of two-particle system is the eigenfunction of the operator
$\hat M_I^2\;(\hat M_I)$, and in the case of pion with constituents of
equal mass has the following form (see, e.g., \cite{KrT23, KrT02}):
$$
\varphi(s(k)) = \sqrt[4]{s}\,k\,u(k)\;,\quad s = 4(k^2 + M^2)\;,
$$
\begin{equation}
	\int\,u^2(k)\,k^2\,dk = 1\;,
	\label{phi}
\end{equation}
where $u(k)$ is a model quark-antiquark wave function and
$M_u = M_{\bar d} = M$ is the mass of the constituents.
The isospin
symmetry is supposed to be valid exactly.

Now we construct the form factors in our version of IF of RQM. Two points
make our version different from the often-used procedure. First, from the
point of view of group theory the corresponding Energy-Momentum Tensor (EMT)
parameterization procedure represents the realization of the analog of the
well-known Wigner -- Eckart theorem on the Poincar\'e group
\cite{KrT05,Edm55}. Second, we consider the form factors as functionals
given by some generalized functions (distributions),  defined on a space
of test functions (see, e.g., \cite{BoL90}, and also \cite{LoC17, KrT02,
KrT03, KrT05}). Thus, in the case of pion the general method of the
relativistic invariant parametrization of matrix elements of local
operators \cite{ChS63} (see also \cite{CoL20}) gives for the EMT matrix
element $T^{(\pi)}_{\mu\nu}(0)$ and electromagnetic current
$j^{(\pi)}_{\mu}(0)$ the following equations:
$$
\langle \vec p_\pi\left|T^{(\pi)}_{\mu\nu}(0)\right|\vec p\,'_\pi\rangle =
\frac{1}{2}G^{(\pi)}_{10}(Q^2)K'_\mu K'_\nu -
$$
\begin{equation}
	- G^{(\pi)}_{60}(Q^2)\left[Q^2g_{\mu\nu} + K_\mu K_\nu\right]\;,
	\label{Tpi}
\end{equation}
\begin{equation}
	\langle\vec p_\pi\left|j^{(\pi)}_{\mu}(0)\right|\vec p\,'_\pi\rangle  =
	K'_\mu\,F^{(\pi)}(Q^2)\;,
	\label{jpi}
\end{equation}
where $G^{(\pi)}_{10}, G^{(\pi)}_{60}$ are the gravitational form factors of the pion,
$g_{\mu\nu}$ is the metric tensor, $F^{(\pi)}$ is electromagnetic pion form factor,
$p_\pi'$ and $p_\pi$ are 4-momenta of pion in initial and final states,
respectively, and
$$
K_\mu = (p_\pi - p_\pi')_\mu\,,\quad K'_\mu = (p_\pi + p_\pi')_\mu \;,
$$
$$
Q^2 = - t = - K_\mu^2\;.
$$
Note that our GFFs in (\ref{Tpi}) are connected with the commonly used
pion GFFs \cite{PoS18, Pag66} by the relations:
\begin{equation}
	A^{(\pi)}(Q^2) = G^{(\pi)}_{10}(Q^2)\;,\;
	D^{(\pi)}(Q^2) = -2\,G^{(\pi)}_{60}(Q^2)\;.
	\label{ADpiG16}
\end{equation}
We present the decompositions of the left-hand side (lhs) of
(\ref{Tpi}), (\ref{jpi})  over the basis  (\ref{Pk}) with the pion quantum
numbers  $J=l=S=0$ as a superposition of the same tensors as in the
right-hand side (rhs) of (\ref{Tpi}), (\ref{jpi}). Thus, we obtain the pion
EFF and GFFs in the following form of the functionals defined on two-quark
wave functions (\ref{phi}):
$$
G^{(\pi)}_{i0}(Q^2) =
\int\,d\sqrt{s}\,d\sqrt{s'}\,
\varphi(s)\tilde G_{i0}(s,Q^2,s')\varphi(s')\;,
$$
\begin{equation}
	i=1,6\;,
	\label{int ds=Gpi}
\end{equation}
\begin{equation}
F^{(\pi)}(Q^2) =	\int\,d\sqrt{s}\,d\sqrt{s'}\,
	\varphi(s)\tilde F(s\,,Q^2\,,s')\, \varphi(s')\;.
	\label{int ds=Fpi}
\end{equation}
Here $\tilde G_{i0}(s,Q^2,s'),\,i=1,6$ and $\tilde F(s\,,Q^2\,,s')$
are the Lorentz-invariant regular distributions.
To construct the invariant distributions in rhs of (\ref{int ds=Gpi}),
(\ref{int ds=Fpi}) we use a modified impulse
approximation (MIA) that we first formulated earlier (see, e.g., Refs.
\cite{KrT02,KrT03} and the review \cite{KrT09}) In contrast to the
baseline impulse approximation, MIA is formulated in terms of the reduced
matrix elements, that is form factors, and not in terms of the operators
themself. So, in MIA there appear important objects -- the free  form
factors describing the characteristics of systems without
interaction. Now we use the gravitational and electromagnetic form
factors of a system of two constituent quarks without interaction in rhs
of (\ref{int ds=Gpi}), (\ref{int ds=Fpi}). Free two-particle form factors
in MIA are given by invariant regular distributions that can be calculated
by the methods of relativistic kinematics. The corresponding matrix
elements in the basis (\ref{Pk}) with the pion quantum numbers have the
following form:
$$
\langle P,\sqrt{s}\left|T^{(0)}_{\mu\nu}(0)\right|P\,',\sqrt{s'}\rangle =
$$
$$
= \frac{1}{2}G^{(0)}_{10}(s,Q^2,s')A'_\mu A'_\nu -
$$
\begin{equation}
	- G^{(0)}_{60}(s,Q^2,s')\left[Q^2\,g_{\mu\nu\nu} + A_\mu A_\nu\right]\;,
	\label{T0}
\end{equation}
$$
\langle\vec P,\sqrt s,\mid j_\mu^{(0)}(0) \mid \vec P',\sqrt{s'}\rangle =
$$
\begin{equation}
	= A'_\mu F^{(0)} (s,Q^2,s')\;,
	\label{j0}
\end{equation}
where $G^{(0)}_{i0}(s,Q^2,s'),\,i=1,6$ are free two-particle GFFs,
$F^{(0)} (s,Q^2,s')$ is the free two-particle EFF, and
$$
A_\mu = \left(P - P'\right)_\mu\;,\quad A^2 = t = -Q^2\;,
$$
$$
A'_\mu = \frac{1}{Q^2}\left[(s - s' + Q^2)P_\mu + (s' - s +
Q^2)P'_\mu\right]\;.
$$
In lhs, zero discrete quantum numbers in state
vectors are ignored.
The Clebsch-Gordan  decomposition of the matrix elements
(\ref{T0}), (\ref{j0}) over the basis of individual spins and momenta
(\ref{p1p2}) gives the expressions for free two-particle form factors in
(\ref{T0}), (\ref{j0}) in terms of one-particle form factors of the
constituents (see for details \cite{KrT23, KrT02}).

These form factors are the functions of the
invariant masses of the two-particle system in the initial and the final
states and depend on the momentun-transfer square as a parameter.
They are the regular generalized functions, the distributions
corresponding to the functionals given by  the two-dimensional
integrals over the invariant masses \cite{KrT02} in the spirit of
the dispersion approach. This means, for instance, that the static limits
at $Q^2\to 0$ are to be understood in a weak sense.

Now we substitute the derived free two-particle form factors instead of
the distributions in (\ref{int ds=Gpi}), (\ref{int ds=Fpi}) taking into
account the quark composition of the neutral pion
$(u\bar u - d\bar d)/\sqrt{2}$. We obtain in MIA the following form for
(\ref{int ds=Gpi}), (\ref{int ds=Fpi}) similar to the expressions
given in \cite{KrT23}:
$$
G^{(\pi)}_{10}(Q^2) =
\frac{1}{4}\sum_{q=u,d}\left[g^{(q)}_{10}(Q^2)+g^{(\bar q)}_{10}(Q^2)\right]\,G^{(\pi)}_{110}(Q^2) +
$$
\begin{equation}
	+ \frac{1}{2}\sum_{q=u,d}\left[g^{(q)}_{40}(Q^2)+g^{(\bar q)}_{40}(Q^2)\right]\,G^{(\pi)}_{140}(Q^2)\;,
	\label{Gpi1}
\end{equation}
$$
G^{(\pi)}_{60}(Q^2) =
\frac{1}{4}\sum_{q=u,d}\left[g^{(q)}_{10}(Q^2)+g^{(\bar q)}_{10}(Q^2)\right]\,G^{(\pi)}_{610}(Q^2) +
$$
$$
+ \frac{1}{2}\sum_{q=u,d}\left[g^{(q)}_{40}(Q^2)+g^{(\bar q)}_{40}(Q^2)\right]\,G^{(\pi)}_{640}(Q^2) +
$$
\begin{equation}
	+  \frac{1}{2}\sum_{q=u,d}\left[g^{(q)}_{60}(Q^2)+g^{(\bar q)}_{60}(Q^2)\right]\,G^{(\pi)}_{660}(Q^2)\;,
	\label{Gpi6}
\end{equation}
$$
F^{(\pi)}(Q^2) =
\frac{1}{2}\sum_{q=u,d}\left[f^{(q)}_{10}(Q^2)+f^{(\bar q)}_{10}(Q^2)\right]\,F^{(\pi)}_{10}(Q^2) +
$$
\begin{equation}
	+ \frac{1}{2}\sum_{q=u,d}\left[f^{(q)}_{30}(Q^2)+f^{(\bar q)}_{30}(Q^2)\right]\,F^{(\pi)}_{30}(Q^2)\;,
	\label{Fpi}
\end{equation}
where $g^{(q)}_{i0}(Q^2)\,,\,q=u,\bar d\,,\,i=1,4,6$ are the GFFs of the
constituent quarks and $f^{(q)}_{k0}(Q^2)\,,\,q=u,\bar d\,,\,k=1,3$ are the EFFs of the
constituent quarks given below.

The form factors in the rhs of the equations (\ref{Gpi1}),
(\ref{Gpi6}) are given now in terms of the integrals \cite{KrT21}.
 Integral representations for electromagnetic and gravitational form factors
of composite systems in our approach are double integrals of special form,
which are analogs of dispersion integrals over the composite-system
mass \cite{TrS69} (see also \cite{AnS87}):
\begin{equation}
	G^{(\pi)}_{ij0}(Q^2) =\int d\sqrt{s} d\sqrt{s'}
	\varphi(s) G^{(0)}_{ij0}(s\,,Q^2\,,s')\varphi(s')\;,
	\label{Gpi1610}
\end{equation}
\begin{equation}
	F^{(\pi)}_{k0}(Q^2) =\int d\sqrt{s} d\sqrt{s'}
	\varphi(s) F^{(0)}_{k0}(s\,,Q^2\,,s')\varphi(s')\;,
	\label{Fpi130}
\end{equation}
where
$i=1,6$; $j=1,4$ for $i=1$ and $j=1,4,6$ for $i=6$; $k
=1,3$; $G^{(0)}_{1i0}(s\,,Q^2\,,s'),$ $\,G^{(0)}_{6j0}(s\,,Q^2\,,s'),$
$F^{(0)}_{k0}(s\,,Q^2\,,s')$ are the components of the free form factors
that describe the system of two free particles with total quantum numbers
of pion, given in Appendix, $\varphi(s)$ is the pion wave function in the
sense of RQM (\ref{phi}), $s'\,,s$ are the invariant masses of the free
two-particle system in the initial and final states, respectively.

Although we do not calculate the mechanical radius here, it seems
necessary to note that in MIA the terms $G^{(\pi)}_{6k0}(Q^2),\,k=1,4$
are singular at zero. A method of regularization has been suggested in
our paper \cite{KrT22}; we have used the fact that in nonrelativistic
limit the corresponding singularity vanishes. Now we obtain
$G^{(\pi)}_{6k0}(Q^2),\,k=1,4$ in the form
$$
G^{(\pi)}_{6k0}(Q^2) =\int d\sqrt{s} d\sqrt{s'}
\varphi(s) G^{(R)}_{6k0}(s\,,Q^2\,,s')\varphi(s')\;,
$$
\begin{equation}
	k=1,4\;.
	\label{Gpi610R}
\end{equation}
The regularized free two-particle form factors
$G^{(R)}_{6k0}(s\,,Q^2\,,s')$ are given in Appendix.

The one-particle form factors of the constituent quarks are connected with
the commonly used ones by the following relations
(the isospin symmetry is supposed to be valid exactly)
\cite{KrT23, KrT02}:
$$
g^{(q)}_{10}(Q^2) =
\frac{1}{\sqrt{1+Q^2/4M^2}}\left[\left(1 +\frac{Q^2}{4M^2}\right)\right.
A^{(q)}(Q^2) -
$$
\begin{equation}
	- \left. 2\frac{Q^2}{4M^2}J^{(q)}(Q^2)\right]\;,
	\label{g10}
 \end{equation}
\begin{equation}
	g^{(q)}_{40}(Q^2) = -\,\frac{1}{M^2}\frac{J^{(q)}(Q^2)}{\sqrt{1 + Q^2/4M^2}}\;,
	\label{g40}
\end{equation}
\begin{equation}
	g^{(q)}_{60}(Q^2) = -\,\frac{1}{2}\sqrt{1 + \frac{Q^2}{4M^2}}D^{(q)}(Q^2)\;,
	\label{g60}
\end{equation}
$$
f^{(q)}_{10}(Q^2) = \frac {1}{\sqrt {1 + Q^2/4M^2}}\,G^{(q)}_E(Q^2)\;,
$$
\begin{equation}
	f^{(q)}_{30}(Q^2) = -\frac {2}{M^2\sqrt {1 + Q^2/4M^2}}\,G^{(q)}_M(Q^2)\;,
	\label{Bal}
\end{equation}
where $A^{(q)},\,J^{(q)},\,D^{(q)}$ are the conventional GFFs of
particles with spin $1/2$ \cite{PoS18, Pag66};  $G^{(q)}_{E,M}$ are
conventional Sachs electrical and magnetic form factors of constituent
quarks (see, e.g., \cite{BaY95}).  We assume that the GFFs and EFFs of
$u$- and $\bar d$-quarks satisfy the following conditions:
$$
g^{(u)}_{i0} = g^{(\bar u)}_{i0} = g^{(d)}_{i0} = g^{(\bar d)}_{i0}\;, i=1,4,6;
$$
\begin{equation}
f^{(q)}_{k0} = -f^{(\bar q)}_{k0}\;, k=1,3\;.
	\label{gfud}
\end{equation}

\section{Results and Discussions}
\label{sec: Sec 4}
\subsection{Neutral pion properties}
\label{sec:Sec4:results}
\subsubsection{Calculation}
\label{sec:Sec4:results:calc}
Now we use the method outlined above to calculate MSRs using quark GFFs
and EFFs (\ref{g10}) - (\ref{Bal}) in the form:
$$
A^{(q)}(Q^2)=f_q(Q^2)\;,\quad J^{(q)}(Q^2) = \frac{1}{2}f_q(Q^2)\;,
$$
$$
	D^{(q)}(Q^2) = D_q\,f_q(Q^2)\;,
$$
\begin{equation}
G^{(q)}_{E}(Q^2) = e_qf_q(Q^2)\;,\quad G^{(q)}_{M}(Q^2) = \mu_q\,f_q(Q^2)\;,
	\label{AJDGfq}
\end{equation}
where $D_q$ is the quark $D$-term, $e_q$ is the charge of the quark,
$\mu_q$ is the quark magnetic moment incorporating its anomalous part.
The functions in rhs of Eqs. (\ref{AJDGfq})
must ensure the standard static limits (see, e.g.,
\cite{PoS18, Gou74}):
$$
A^{(q)}(0)=1\;,\quad J^{(q)}(0) = \frac{1}{2}\;,\quad D^{(q)}(0) = D_q\;,
$$
\begin{equation}
G^{(q)}_{E}(0) =  e_q\;,\quad G^{(q)}_{M}(0) = \mu_q\;.
	\label{Ag0}
\end{equation}
The function $f_q(Q^2)$ in (\ref{AJDGfq}) is chosen in the form
that we have used in \cite{KrT22, KrT09, KrT01} when we calculated the
gravitational and electromagnetic properties of charged pions,
\begin{equation}
	f_q(Q^2) = f_{\bar q}(Q^2) =
	\frac{1}{1 + \ln\left(1 + \langle r^2_q\rangle Q^2/6\right)}\;.
	\label{fq}
\end{equation}
The constituent-quark mass radius $\langle r^2_q\rangle$ is assumed to be
equal to its charge radius and defined by the relation
(see Refs. \cite{VoL90, PoH90, TrT94, CaG96}):
\begin{equation}
	\langle r^2_q\rangle  \simeq \frac{0.3}{M^2}\;.
	\label{rq}
\end{equation}
We have used these assumptions earlier in \cite{KrT09, KrT01} and \cite{KrT22}.

In the calculation, we use three kinds of model two-quark wave
function with parameters $b$ (quoted in Table \ref{tab:table2}) fixed in
our previous works \cite{KrT09, KrT01, Kru97}:
\begin{equation}
	u(k) = 2\left(1/(\sqrt{\pi}\,b^3)\right)^{1/2}\exp\left(-\,k^2/(2\,b^2)\right)\;,
	\label{wfHO}
\end{equation}
\begin{equation}
	u(k) = 16\left(2/(7\pi\,b^3)\right)^{1/2}\left(1 + k^2/b^2\right)^{-3}\;,
	\label{wfPL3}
\end{equation}
\begin{equation}
	u(k) = 4\left(2/(\pi\,b^3)\right)^{1/2}\left(1 + k^2/b^2\right)^{-2}\;.
	\label{wfPL2}
\end{equation}
The quark mass is the same as in our papers on EFFs of charged pions
(see, e.g., \cite{KrT01}), $M$= 0.22 GeV.

\subsubsection{Results: the electromagnetic form factor}
\label{sec:Sec4:results:EFF}
We begin our discussion of the results with the case of the EFFs of
neutral pion (\ref{Fpi}). It is seen from the expressions
(\ref{Fpi130}) and (\ref{gfud}), that the charge form factor of neutral
pion is identically zero for any interaction model (\ref{wfHO}) -
(\ref{wfPL2}). This means that our instant form relativistic impulse
approximation approach automatically
satisfies the requirement of the $C$-symmetry discussed in
Sec.~\ref{sec:Sec2:charge}.

\subsubsection{Results: the $A$ form factor and the mass radius}
\label{sec:Sec4:results:mass}

The results of the calculation of the mass and energy MSRs using 
wave functions
(\ref{wfHO}) - (\ref{wfPL2}) are given in  Table \ref{tab:table2}.
\begin{table}
	\caption{\label{tab:table2} The results of calculating the
		static gravitational moments of the $\pi^0$ with different model
		wave functions (\ref{wfHO}) - (\ref{wfPL2}). The mass of
		constituent quarks is $M = 0.22$ GeV, their MSR is given in
		(\ref{rq}) and the $D$-term $D_q$ is taken from (\ref{Dqmech}).
		The deviation from the given values of
		$S^{(\pi)}_D$ when the parameter $D_q$ is
		varied in the interval (\ref{Dqmech}) is about 0.1\%.}
	\begin{tabular}{lccc}
		\hline
		\hline
		Model                                 &(\ref{wfHO})     &(\ref{wfPL3})    &(\ref{wfPL2})  \\
		%\hline
		$b$, GeV                              &0.3500           &0.6131           &0.4060          \\
		\hline
		$A^{(\pi)}\,'(0)$, GeV$^{-2}$         &1.074            &1.134            &1.217           \\
		$\langle r^2\rangle^{1/2}_{\rm mass}$, fm &0.50             &0.52             &0.53            \\
		$-D^{(\pi)}(0)$                       &0.704--0.700     &0.653--0.649     &0.620--0.617    \\
		$-D^{(\pi)}\,'(0)$,GeV$^{-2}$         &2.331--2.327     &1.978--1.976     &1.773--1.771    \\
		$S^{(\pi)}_D$, fm                     &0.88             & 0.84            &0.82            \\
		$\langle r^2\rangle^{1/2}_{\rm energy}$, fm &0.922--0.928&0.847--0.854&0.795--0.800         \\
		\hline
		\hline
	\end{tabular}
\end{table}
Only the parameters used previously for obtaining the charge MSR
enter the equations (\ref{ADpiG16}), (\ref{Gpi1}),
(\ref{Gpi1610}) and (\ref{R2mass}) used to find $A^{(\pi)}(Q^2)$.
The quark $D$-term $D_q$
(\ref{Ag0}), the new parameter, is only exploited below to obtain the
form factor $D^{(\pi)}(Q^2)$.

\subsubsection{Results: the $D$ form factor and its slope $S^{(\pi)}_D$}
\label{sec:Sec4:results:D}
\begin{figure}[h!]
        \epsfxsize=0.9\textwidth
        \centerline{\psfig{figure=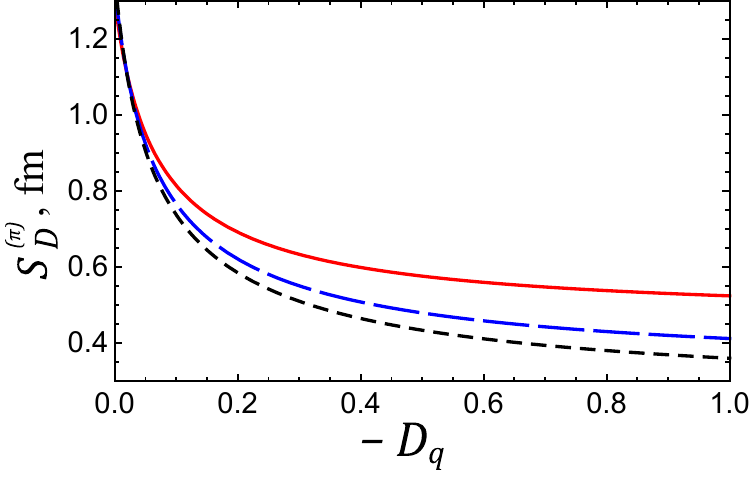,width=9cm}}
        \vspace{0.3cm}
        \caption{The  slope at zero of the normalized to pion $D$-term
                form factor $D$ of pion as a function of the $D$-term of
                the constituent quark $D_q$ calculated with various model
                wave functions (\ref{wfHO})  -- (\ref{wfPL2}). Full line
                (red) -- with wave function (\ref{wfHO}); dashed  line
                (blue) -- with (\ref{wfPL3}), short-dashed line (black) --
                with (\ref{wfPL2}).}
        \label{fig:1}
\end{figure}

As we mentioned above, in the paper \cite{KuS18} the first value of
(\ref{R2mech2}) was extracted, using a phenomenological approach, from the
data on the reaction $\gamma^*\gamma\to \pi^0\pi^0$,
\begin{equation}
        S^{(\pi)}_D = \left(0.82 - 0.88\right)\,\hbox{fm}\;.
        \label{SDkus}
\end{equation}
In the present paper we use this result only to fix the quark $D$-term
$D_q$, that enters the formulas (\ref{ADpiG16}),
(\ref{Gpi6}), (\ref{Gpi610R}), (\ref{g60}), (\ref{AJDGfq}) for GFF
$D^{(\pi)}(Q^2)$. As all other parameters have been fixed previously
\cite{KrT01, Kru97}, now all calculated values are the functions of $D_q$.
The dependence of the normalized slope (\ref{R2mech2}), $S^{(\pi)}_D$, on
$D_q$, calculated with three model wave functions
(\ref{wfHO}) - (\ref{wfPL2}), is given in Fig. \ref{fig:1}.
It demonstrates that, for each of the model
wave functions, there is an interval of the values of $D_q$ which gives the
interval (\ref{SDkus}) of the values of $S^{(\pi)}_D$. Moreover, there
exists the interval of the variable $D_{q}$ for which $S^{(\pi)}_D\,$
falls in  the interval (\ref{SDkus}) for all wave functions
(\ref{wfHO})--(\ref{wfPL2}) simultaneously, without any additional
variation of parameters:
\begin{equation}
        D_q = -\left(0.0715 - 0.0709\right)\;.
        \label{Dqmech}
\end{equation}
Similar result was obtained for electroweak properties of pion in RQM
\cite{KrT01}, when the quark parameters (mass, charge MSR, anomalous
magnetic moment) were fixed simultaneously in model functions
(\ref{wfHO})--(\ref{wfPL2}). Now we add the quark $D$-term to the
previous set and obtain the common approach to the description of
pion EFFs and GFFs.

The results of calculation of gravitational characteristics of neutral
pion are given in Table \ref{tab:table2}. The values connected solely with
form factor $A$, (\ref{ADpiG16}), (\ref{Gpi1}) can be considered as a kind
of predictions because they do not depend on the "new"$\,$ parameter $D_q$.
The condition $A^{(\pi)}(0) = 1$ is satisfied automatically
for any values of parameters if the relation (\ref{Ag0}) for quark form
factor is fulfilled. It is important to note that the deviation of the
values of $S^{(\pi)}_D$ from values given in Table \ref{tab:table2}
when $D_q$ is changed in the interval (\ref{Dqmech}) is approximately
0.1\%.

\begin{figure}[h!]
        \epsfxsize=0.9\textwidth
        \centerline{\psfig{figure=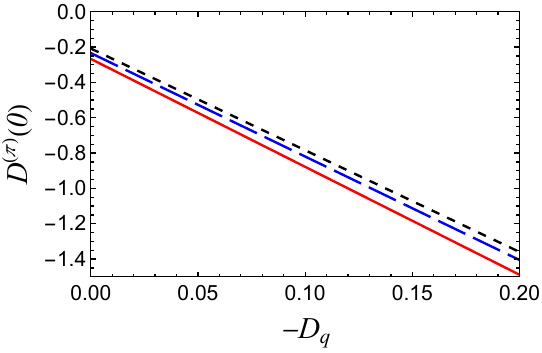,width=9cm}}
        \vspace{0.3cm}
        \caption{$D$-term $D^{(\pi)}(0)$ of neutral pion
                as a function of the $D$-term of
                the constituent quark $D_q$ calculated with various model
                wave functions (\ref{wfHO})  -- (\ref{wfPL2}). Legend is the
                same as in Fig. \ref{fig:1}.}
        \label{fig:2}
\end{figure}

One should note that only the quark
components were taken into account when extracting (\ref{SDkus}) from
the experiment in the paper \cite{KuS18}; gluon components increase
the total contribution to the slope \cite{HaO23}. Thus, it is possible to
consider the value $S^{(\pi)}_D$ (\ref{SDkus}) as a kind of a lower
bound. An increase in $S^{(\pi)}_D$ will shift (see Fig. \ref{fig:1}) the
interval (\ref{Dqmech}) towards smaller absolute values. This in turn
will lead to an increase of the values $D^{(\pi)}(0)$ as compared to the
values given in Table \ref{tab:table2} (See Fig. \ref{fig:2}).
Note that the value (\ref{SDkus}) in \cite{KuS18} is quite close to the
result of calculation in \cite{XuD24}, $S^{(\pi)}_D=$ 0.81 fm.

\subsection{Comparison with other approaches}
\label{sec:Sec4:comparison}
Table~\ref{tab:table1} presents our value of the mass MSR together with
those obtained within several different approaches. We see that our value
is somewhat larger than some of the others.
\begin{table}
        \caption{\label{tab:table1} The mean-square mass radius of
                neutral pion (\ref{R2mass}) in different approaches.
                Our result is pre-averaged over the models
                (\ref{wfHO}) - (\ref{wfPL2})and uses the constituent-quark
                mass $M = 0.22$ GeV and the MSR radius defined by
                (\ref{rq}).}
\begin{tabular}{ccc}
\hline
\hline
$\langle r^2\rangle^{1/2}_{\rm mass}$, fm & Reference   & Approach  \\
\hline
        0.517                         &This work    &instant form RQM\\
\hline
        0.52(2)                       &\cite{XuR23} & empirical \\
        0.47                          &\cite{XuD24} & Schwinger function\\
        0.41$\pm$0.01                 &\cite{HaO23} & lattice   \\
        0.32 - 0.39                   &\cite{KuS18} &$\gamma^*\gamma\to\pi^0\pi^0$\\
        $\approx$  0.39               &\cite{LiV24} &holography \\
        0.31$\pm$0.04                 &\cite{AlB21} &sum rules\\
        $\approx$ 0.27                &\cite{FrC19} &Nambu-Jona-Lasinio\\
        0.244                         &\cite{SuX24} & contact
interactions \\
\hline
\hline
\end{tabular}
\end{table}
Possible reasons for this difference
may be understood by recalling the differences in definitions and
methodologies mentioned above in Sec.~\ref{sec: Sec 2}, which we discuss
in more detail here. Indeed, our approach
deals with the degrees of freedom connected with constituent quarks.
Introducing parameters describing the constituents structure we
effectively take into account the gluon degrees of freedom of QCD and
those of the sea quarks. On the other side, the mass MSR, extracted in
a phenomenological framework from the data on the process
$\gamma^*\gamma\to\pi^0\pi^0$ \cite{KuS18}, does not contain gluon
contribution to pion GFF and is considerably smaller than our value.
However, QCD lattice model calculations \cite{HaO23} show that the additive
contribution of the gluon component has the same slope as the quark
component in \cite{KuS18}, and is large. Thus, the account of the gluon
component should increase the value of mass MSR of pion and should bring
the two results closer to each other. Unfortunately, lattice calculations
at low momentum transfers are performed so far at a small number of points
(for $Q^2<$0.25$\,\hbox{GeV}^2$ only at five points) and with rather large
uncertainties; note, moreover, that the mean value of $A^{(\pi)}(0)$ is
slightly greater than unity, cf.\ (\ref{Ag0}). All this might
result in a significant inaccuracy in the calculated slope of
$A^{(\pi)}(Q^2)$ at zero. In the lattice model calculation in Ref.
\cite{HaO23}, the pion mass was taken $m_\pi\approx 170$ MeV, that differs
from the observed value by approximately 25\%. This has no fundamental
importance for the calculation of $A^{(\pi)}(Q^2)$ at finite $Q^2$, but
can be crucial for the slope at zero. Note that in the lattice model
calculations in \cite{Pef23} it was shown that the gluon contribution to
the slope of $A^{(\pi)}(Q^2)$ increases with decreasing the pion mass.
This means that lattice calculations using the pion mass that is closer to
the experimental value should enlarge the slope and diminish the
difference with our result. Therefore, a detailed
additional study is required for a reasonable comparison of the values of
pion mass MSR (\ref{R2mass}) in different approaches.

Let us consider now the numerical values of the slope  $S^{(\pi)}_D$ in
different approaches. Lattice QCD approach with account of gluon
contribution gives \cite{HaO23}:
\begin{equation}
        S^{(\pi)}_D = 0.61\pm 0.07\,\hbox{fm}\;.
        \label{SDlatt}
\end{equation}
Approach in holographic light-front QCD gives the value \cite{LiV24}:
\begin{equation}
	S^{(\pi)}_D \approx 0.60\,\hbox{fm}\;.
	\label{SDhol}
\end{equation}
In the work \cite{LiV24}, the gluon degrees of freedom are incorporated
by account in the Fock space not only of valent quarks but also of
glueballs.

Although the numerical values (\ref{SDlatt}) and (\ref{SDhol}) are smaller
than (\ref{SDkus}), it is necessary to examine the reliability of the
approaches and to recall the criticisms given above.
Firstly, (\ref{SDlatt}) is obtained using the monopole
fitting (see (\ref{Ddip}) for $n$=1) of the pion GFF at $Q^2<$2 GeV$^2$.
However, the slope at zero is usually calculated exploiting the linear
fitting at small $Q^2$ (see, e.g., \cite{Ame84}). Such fitting using the
points obtained in \cite{HaO23} for $Q^2<$0.25 GeV$^2$ (so far,
unfortunately, only four points) gives $D^{(\pi)}(0)\approx$ -0.81 and
$S^{(\pi)}_D\approx$ 0.41 fm with errors which overlap the value
(\ref{SDkus}) obtained in \cite{KuS18}.

When discussing the reliability of the monopole fitting for
$D^{(\pi)}(Q^2)$ at low $Q^2$, it is necessary to take into account
that in the case of the nucleon form  factor $D$, the contribution of
the gluon component is suppressed as compared to quark contribution, as
it was shown in \cite{Pef23}. This means that the contributions of gluons
and of quarks are principally qualitatively of different structure, thus
making invalid even the functional form (\ref{Ddip}) for low momentum
transfers and also its corollaries (including (\ref{SDlatt})) for pion.
So, additional lattice calculations at low $Q^2$ are requiered. Secondly,
in the paper \cite{HaO23}, the quark mass was used, which corresponds to
the value of pion mass that exceeds the observable value approximately by
25\%, comparable to the difference between (\ref{SDkus}) and
(\ref{SDlatt}). In this connection we note, that the authors of the paper
\cite{Pef23} obtained a decrease of gluon contribution to $D^{(\pi)}(Q^2)$
with decreasing pion mass.

As to deriving (\ref{SDhol}) in \cite{LiV24}, the parameters defining the
weight of states of scalar glueballs were obtained under the
assumption of chiral symmetry which is obviously broken. It follows
that in the case of more realistic approximations (\ref{SDhol})
may change its value.

As can be seen from the above, a number of the results for the mass MSR
given in Table~\ref{tab:table1} are likely to be modified by using more adequate
approximations and performing more accurate calculations. There are
indications that these results will be closer to those of our approach
in this case. Recall that our result is obtained without free parameters
and is a strong prediction of our approach. So, further
theoretical study of a pion GFF $D^{(\pi)}(Q^2)$, in particular, by
lattice calculations, is wellcome.
Our opinion is that, despite the fact that the results obtained in two
different approaches (\ref{SDlatt}) and (\ref{SDhol}) are close to one
another, the discussion above supports our assumption that extracted in
\cite{KuS18} from the experiment value  (\ref{SDkus})
presents the lower bound of $S^{(\pi)}_D$ .

\section{Conclusions}
\label{sec: Sec 5}

In this paper we continue the study of the pion structure using our
instant-form relativistic impulse approximation approach to composite
particle theory.
We use a
version of the instant-form of Dirac relativistic quantum mechanics
complemented by an essentially
relativistic variant of impulse approximation, formulated previously
in our papers on electroweak structure of composite particles. The
calculations were carried out in three model of interaction of constituent
quarks with different types of confinement, that is with three types  of
quark-antiquark wave functions in pion. The most significant results of
the paper are obtained using only the set of parameters which has been
derived in our old papers on pion electroweak structure, without any new
variation of them.

We believe that two main results of the paper are: the size
and the EFF of the neutral pion.

1) Although the positive definiteness of the normal force
connected with the stability of a particle
is a suitable physical criterion to define the size of a particle in term
of largely used mechanical radius, we choose the mass radius. We give a
detailed argumentation for our choice. The calculated value of mass MSR is
$(0.5-0.53)$ fm, the spreading being due to three model quark-antiquark
wave functions.

2) The calculated in the present work charge form factor of the neutral
pion is identically zero. This means that our essentially  relativistic
nonperturbative approach automatically gives the result which coincides
with the general statement that single-photon coupling is forbidden by
charge-conjugation invariance.

We compare our results for different MSRs with the results of other
approaches. To make such a comparison  complete, we
consider pion form factor $D$, too. For its calculation we add one  more
parameter, the quark $D$-term that we obtain by fitting the slope of
the pion form factor $D$ at zero momentum transfer to a chosen
experimental value.
The careful detailed analyzis, qualitative as well as quantitative, of the
results of calculations of a set of MRSs in different approaches is
presented in the paper. We show that for the deeper understanding of the
problem of pion size in some approaches, including lattice QCD
calculations, the additional further study is important,
which should properly take into account the
asymptotic
behavior and of the relation between quark and gluon degrees of
freedom.

A number of expressions derived in the present work can be exploited for
the calculation of electromagnetic and gravitational properties of other
composite hadrons.

\section*{Appendix. Free two-particle gravitational and electromagnetic
form factors}

GFFs of two noninteracting fermions with spin 1/2 (\ref{Gpi1610}):
$$
G^{(0)}_{110}(s, Q^2, s') = -\,\frac{R(s, Q^2, s')\,Q^2}{\lambda(s,-Q^2,s')}
$$
$$
\times\left[(4M^2+Q^2)\lambda(s,-Q^2,s') - \right.
$$
$$
- \left.3\,Q^2(s + s'+ Q^2)^2\right]\cos(\omega_1+\omega_2)
\;,
\eqno{(A1)}
$$
$$
G^{(0)}_{140}(s, Q^2, s') = -3\,M\,\frac{R(s, Q^2, s')\,Q^4}{\lambda(s,-Q^2,s')}
$$
$$
\times\xi(s,Q^2,s')(s+s'+Q^2)\sin(\omega_1+\omega_2)\;,
\eqno{(A2)}
$$
$$
G^{(0)}_{610}(s, Q^2, s') = \frac{1}{2}\,R(s, Q^2, s')
$$
$$
\times\left[(s + s' + Q^2)^2 - \right.
$$
$$
- \left.(4M^2 + Q^2)\lambda(s,-Q^2,s')/Q^2\right]\cos(\omega_1+\omega_2)
\;,
\eqno{(A3)}
$$
$$
G^{(0)}_{640}(s, Q^2, s') = -\,\frac{M}{2}\,R(s, Q^2, s')
$$
$$
\times\xi(s,Q^2,s')(s+s'+Q^2)\sin(\omega_1+\omega_2)\;,
\eqno{(A4)}
$$
$$
G^{(0)}_{660}(s, Q^2, s') = R(s, Q^2, s')
$$
$$
\times\lambda(s,-Q^2,s')\cos(\omega_1+\omega_2)\;.
\eqno{(A5)}
$$

EFF of two noninteracting fermions with spin 1/2 (\ref{Fpi130}):

$$
F^{(0)}_{10}(s\,,Q^2\,,s') =
R(s, Q^2, s')Q^2
$$
$$
\times(s+s'+Q^2)\cos(\omega_1+\omega_2)\;,
\eqno{(A6)}
$$
$$
F^{(0)}_{30}(s\,,Q^2\,,s') = - R(s, Q^2, s')Q^2
$$
$$
\times\frac{M}{2}\xi(s,Q^2,s')\sin(\omega_1 + \omega_2)\;,
\eqno{(A7)}
$$
where
$$
R(s, Q^2, s') = \frac{(s + s'+Q^2)}{2\sqrt{(s-4M^2) (s'-4M^2)}}\,
$$
$$
\times\frac{\vartheta(s,Q^2,s')}{{[\lambda(s,-Q^2,s')]}^{3/2}}\;,
$$
$$
\xi(s,Q^2,s')=\sqrt{-(M^2\lambda(s,-Q^2,s')-ss'Q^2)}\;,
$$
$\omega_1$ and $\omega_2$ are the Wigner spin-rotation parameters:
$$
\omega_1 =
\arctan\frac{\xi(s,Q^2,s')}{M\left[(\sqrt{s}+\sqrt{s'})^2 + Q^2\right] + \sqrt{ss'}(\sqrt{s} +\sqrt{s'})}\;,
$$
$$
\omega_2 = \arctan\frac{ \alpha (s,s') \xi(s,Q^2,s')} {M(s + s' + Q^2) \alpha (s,s') + \sqrt{ss'}(4M^2 + Q^2)}\;,
$$
$\alpha (s,s') = 2M + \sqrt{s} + \sqrt{s'} $,
$\vartheta(s,Q^2,s')= \theta(s'-s_1)-\theta(s'-s_2)$,
$\theta$ is the Heaviside function.
$$
s_{1,2}=2M^2+\frac{1}{2M^2} (2M^2+Q^2)(s-2M^2)
$$
$$
\mp \frac{1}{2M^2} \sqrt{Q^2(4M^2+Q^2)s(s-4M^2)}\;,
$$
$$
\lambda(a,b,c) = a^2 + b^2 +c^2 - 2(ab + ac + bc)\;,
$$
$M$ is the mass of constituent quarks.

GFFs of two noninteracting fermions with spin 1/2
in minimal extension of MIA (\ref{Gpi610R}):
$$
G^{(R)}_{610}(s, Q^2, s') = -\frac{1}{2}\,\tilde R(s, Q^2, s')
$$
$$
	\times 4M^2\,Q^2\cos(\omega_1+\omega_2)\;,
\eqno{(A8)}
$$
$$
G^{(R)}_{640}(s, Q^2, s') = -\,\frac{M}{2}\,\tilde R(s, Q^2, s')
$$
$$
	\times 8M^2\tilde\xi(s,Q^2,s')\sin(\omega_1+\omega_2)\;,
	\eqno{(A9)}
$$
here
$$
\tilde\xi(s,Q^2,s')=
$$
$$
	= M\sqrt{\left[(s'-4M^2)-\tilde s_1\right]\left[\tilde s_2 - (s'- 4M^2)\right]}\;,
$$
$$
\tilde s_1 = \left(\sqrt{s-4M^2} - \sqrt{Q^2}\right)^2\;,
$$
$$
	\tilde s_2 = \left(\sqrt{s-4M^2} + \sqrt{Q^2}\right)^2\;.
$$
The functions
$\tilde R(s, t, s')$ and  $R(s, t, s')$ (in ({\it A1}) - ({\it A9}))
differ in cutting function:
$$
	\vartheta(s,t,s')\;\to\;\tilde\vartheta(s,Q^2,s')= \theta(s'-\tilde s_1)-\theta(s'-\tilde s_2)\;.
$$

\end{document}